\documentclass[lnicst]{svmultln}
\usepackage{graphicx}
\usepackage{epstopdf}
\usepackage{makeidx}  % allows for indexgeneration
\begin{document}
\mainmatter              % start of the contribution
\title{Radio Link Planning made easy with a Telegram Bot.}
\titlerunning{Radio Link Planning made easy with a Telegram Bot}  % abbreviated title (for running head)
\author{Marco Zennaro \inst{2} \and Marco Rainone \inst{1}  \and Ermanno Pietrosemoli \inst{2}}

\tocauthor{IMarco Rainone, Marco Zennaro and Ermanno Pietrosemoli}
\institute{SolviTech, Udine, Italy\\
%\email{I.Ekeland@princeton.edu},\\ WWW home page:
%\texttt{http://users/\homedir iekeland/web/welcome.html}
\and
ICTP
Telecommunication/ICT4D Lab\\
Strada Costiera 11, Trieste, Italy}

\maketitle              % typeset the title of the contribution
% \index{Ekeland, Ivar} % entries for the author index
% \index{Temam, Roger}  % of the whole volume
% \index{Dean, Jeffrey}

\begin{abstract}        % give a summary of your paper
Traditional radio planning tools present a steep learning curve. We present BotRf, a Telegram Bot that facilitates the process by guiding non-experts in assessing the feasibility of radio links. Built on open source tools, BotRf can run on any smartphone or PC running Telegram. Using it on a smartphone has the added value that the Bot can leverage the internal GPS to enter coordinates. BotRf can be used in environments with low bandwidth as the generated data traffic is quite limited. We present examples of its use in Venezuela.
%                         please supply keywords within your abstract
\end{abstract}
\section{Introduction}
After many years of teaching wireless networking to people from different backgrounds, we still miss an easy to use, open source tool that can be used both as a simulation for the planning of wireless installations and also as a teaching aid that helps navigate the intricacies of radio frequency propagation, especially arcane to people who have a background in computer sciences but no radio frequency experience. Taking advantages of the facilities of the bot technology, we developed a tool that does all the processing in the remote server, minimizes the amount of data that the user needs to input as well as the traffic that traverses the net, making it particularly suitable for countries with slow or expensive Internet access and for community networks. BotRf is platform independent, runs in Telegram and therefore can be used in any smartphone or even a laptop or desktop. 
It provides simulation of the terrain profile for different refraction index values, simulates the path loss in a wireless point to point link and shows the result in easy to grasp graphics and a numerical value of the link margin in dB.
For point to multipoint applications, BotRf will provide a graph of simulated signal levels available around the transmitting antenna and the margin above the receiver sensitivity.

Community wireless network have sprouted in many countries, both rich and less rich, addressing needs that are not satisfied by commercial operators, either due to  the lack of a high enough return on investment or  because commercial operators  were not well suited to address social aspects unrelated to profits.
We have been conducting training in wireless networking since 1992 in many countries, and striving to make these efforts more effective we first produced the book "Wireless Networking in Developing Countries", freely downloadable from \cite{wndw}, which has been translated into other six languages and is being widely used.
One aspect that is paramount in the planning of wireless links is the determination of the attenuation introduced by the terrain between the transmitter and the receiver, which ultimately determines the feasibility of a given link.
There are many commercial programs meant to solve this problem, most of them making use of digital elevation maps, some are quite costly and others restrict their usage to the radios and antennas of a particular manufacturer.

Splat! Is an open source Linux based program that uses two different models to simulate the attenuation between the transmitter and the receiver \cite{qsl}, and our work uses part of its code to provide a user friendly tool that can benefit both a beginner as well as an advanced user. In particular, BotRf will use SRTM1 digital elevation maps (DEM) \cite{srtm} , which have a resolution of 30 m, instead of the 90 m resolution used by splat!. Furthermore all the processing is done in the server, no computer is required (although one can be used to take advantage of the bigger screen), only smart phones and tablets which are available to more people than the ones who use computers. Very little Internet traffic is used by BotRf, which makes it very fast and inexpensive to use. The built in GPS present in most smart phones can be used to capture directly the latitude and longitude of a place of interest, without the need to manually enter them, thus reducing the data entry errors. The elevation of the site is determined from the DEMs in the server, which is normally more accurate than the elevation measured by the GPS. We have drawn from  our experience of many years of teaching wireless networking to people of different backgrounds, to address the more relevant issues we have encountered. A built in converter allows the often confusing task of calculating dBm from miliwatts and viceversa, as well as other often needed conversions. Results are presented in  easy to understand graphics annotated with the most relevant numerical data.
Telegram\cite{teg} is an open source messaging platform which supports bots \cite{bot}

\section{Radio Frequency Propagation Planning}

There are many  RF planning tools, which can be categorised as Commercial, Vendor Specific, Free or Open. Some  commercial tools use proprietary models and digital elevation maps of varying degrees of resolution. They are quite expensive, require downloading and installation in a specific computer and are not well suited to the needs of developing countries.  No simulation tool is ever 100\% accurate, so its results must always be interpreted and confronted  with additional information pertaining a particular case.
Vendors like Motorola, Ubiquiti Networks, Cambium and Mimosa  (among many others), offer radio propagation planning tools free of charge. They are usually associated to the specific of their radios and antennas offering, often require registration in their site and in general are quite cumbersome to use.

\subsection{Free RF planning tools}

Several free radio frequency planning tools are available. In our trainings and deployments we have used extensively the excellent Radio Mobile program developed by Roger Coude \cite{rm}. It provides very valuable information, but it is built for the Windows operating system only, and requires considerable effort to install and to learn how to use it. There is also an on-line version, which is web based and easier to use, but works only for the frequencies allotted to the radio amateur service and does not specify the amount of first Fresnel zone clearance \cite{rmol} 

"Hey whats that path" profiler is another free on line tool that can quickly draw terrain profiles between two or more points. Does not calculate attenuation, but allows to  change the frequency and the atmospheric refraction index and show how this affects the radio beam \cite{hey}.

The Communications Research Centre Canada offers the Radio Coverage Prediction using the Longley-Rice model, but it is only for point to area, will not work for point to point links and its coverage is limited regions up to 5 degrees in latitude and 10 degrees in longitude {crc}

Converseley, the Polish National Institute of Telecommunications developed PIAST (Platform IT for an Analysis of Systems in Telecommunications) {piast}, a web based tool for point to point RF links planning that will provide a path profile and estimates of the attenuation, using the ASTER GDEM v2 maps.

\subsection{Open RF planning tools}
John Magliacane wrote Splat! (Signal Propagation, Loss, And Terrain Analysis), an open source Linux based program  (offered under the GNU license) that uses both the LR ITM and the Irregular Terrain with Obstructions (ITWOM)  models to simulate the attenuation between the transmitter and the receiver as well as to estimate the point to area coverage of a transmitter for the frequencies between 20 MHz and 20 GHz. splat! has been ported to Windows and is also the basis of QRadioPredict, an experimental software for VHF-UHF propagation prediction and radio coverage analysis. It currently  works on Linux and Windows and has many  additional features \cite{qrp}. 
BotRf modifies the splat! code to work directly with the native file format of the high resolution digital elevation maps and to provide a very user friendly and platform independent user interface that consumes very little bandwidth.

\subsection{ RF Propagation Models}
In \cite{cal}, a thorough analysis of 30 propagation models used for the estimation of path loss is performed, reaching the conclusion that there is not a single one that can offer the best results in every case. Some models make use of digital elevation maps and ray tracing techniques to estimate the path loss, while other use a completely empirical approach, based on many measurements performed in a variety of environments at different frequencies. Then there is the hybrid approach, that combines terrain information with statistical inference from field measurements. They end up stating that "..the landscape of path loss models is precarious: typical best-case performance accuracy of these models is on the order of 12-15 dB root mean square error (RMSE) and in practice it can be much worse". Therefore, in our work we stick with the well known Longley-Rice Irregular Terrain Model ( L-R ITM)   [10], which has been widely used since its original publication in 1968. The advantage of the L-R ITM is that by making use of the terrain profile between transmitter and receiver (derived from digital elevation maps), it will readily detect any obstruction that might preclude the link. So if the model suggests that a links is infeasible, we accept this result and search for an alternative such as raising the antenna height or look for an alternate location. On the other hand, when the model finds a path unobstructed, we take this result with a grain of salt and declare it a "maybe" until further information is available. In particular, the uncertainty is significantly more when using SRTM3 DEM that have a significant chance of ignoring obstacles that are less than 90 m in size, so we will use SRTM1 DEMs in which the resolution is three times better. These strategy has been quite successful over a number of links that we have simulated and later deployed. The path attenuation result of the L-R ITM is calculated by adding to the free space loss the additional losses caused by diffraction and dispersion, as well as statistic losses that have been estimated based on many empirical measurements at different frequencies and over different types of terrain.

Of course a very important factor that must be borne in mind when dealing with RF propagation is that refraction index of the atmosphere causes a bending of the radio beam. This bending normally causes the radio horizon to extend 4/3 beyond the optical horizon, and this is  accounted for by multiplying the real radius of curvature of the earth by a K=4/3 factor and then draw the radio beam as a straight line.
Unfortunately, the atmosphere refraction index can change unexpectedly as it is affected by factors like temperature, humidity, water vapor content and so on, so the K factor might adopt temporarily other values. The  International Telecommunications Union (ITU) \cite{itu} recommends that for critical links in which a very high reliability is required, the obstacles of the terrain must be cleared even when a K less than 1 for distances below 100 km.  For K less than1 the effective radius of the earth is smaller than the real radius, that is the radio horizon is closer that the optical horizon. BotRf allows for easy playing with different K factors and antenna heights to assess links feasibility.
One must also keep in mind that there are also other factors that limit the reliability of the network, for instance traffic overloading and so on, therefore for the case of Internet access it is seldom justified to consider these very special cases of  refraction index vagaries. 
Lastly, and more importantly, we must also note that electromagnetic waves occupy a volume in space  called the Fresnel ellipsoid, extending from the transmitter to the receiver, with maximum girth at midpoint between the two.  The radius of the Fresnel ellipsoid is called the Fresnel zone radius and it grows with the distance from  the transmitter up to the midpoint and then starts diminishing again. This radius is proportional to the wavelength, and since the wavelength of light is around 1 micrometer, the Fresnel zone is negligible for visible light and the optical line of sight (LOS) is essentially a straight line between transmitter and receiver. Radio waves, on the other hand, can have quite big wavelengths, and in order to capture most of the power contained in them, at least 60\% of the Fresnel zone must be cleared. The radius of the Fresnel zone in meters at a point d1 km from the transmitter and d2 from the receiver, for a wavelength of wl meters is given by:

\

$F1={(wl*d1*d2)\over(d1+d2)}$

\

In a wireless link, we try to avoid that any obstacle protrudes more than 40\% in the Fresnel zone, this means that the antennas must be raised above the minimum required to draw a line from the transmitter to the receiver (the optical LOS).
In the following graph, the blue line represents the optical LOS, while the magenta curve is the envelope of Fresnel zone. Note that the terrain is cleared by the LOS, but the lower part of the Fresnel zone is obstructed. This is known as the grazing condition and implies a loss of 6 dB in the received signal power (we are only getting 25\% of the power) as shown in Fig. 1.

\begin{figure}[h]
\centering
\includegraphics[width=4.5in]{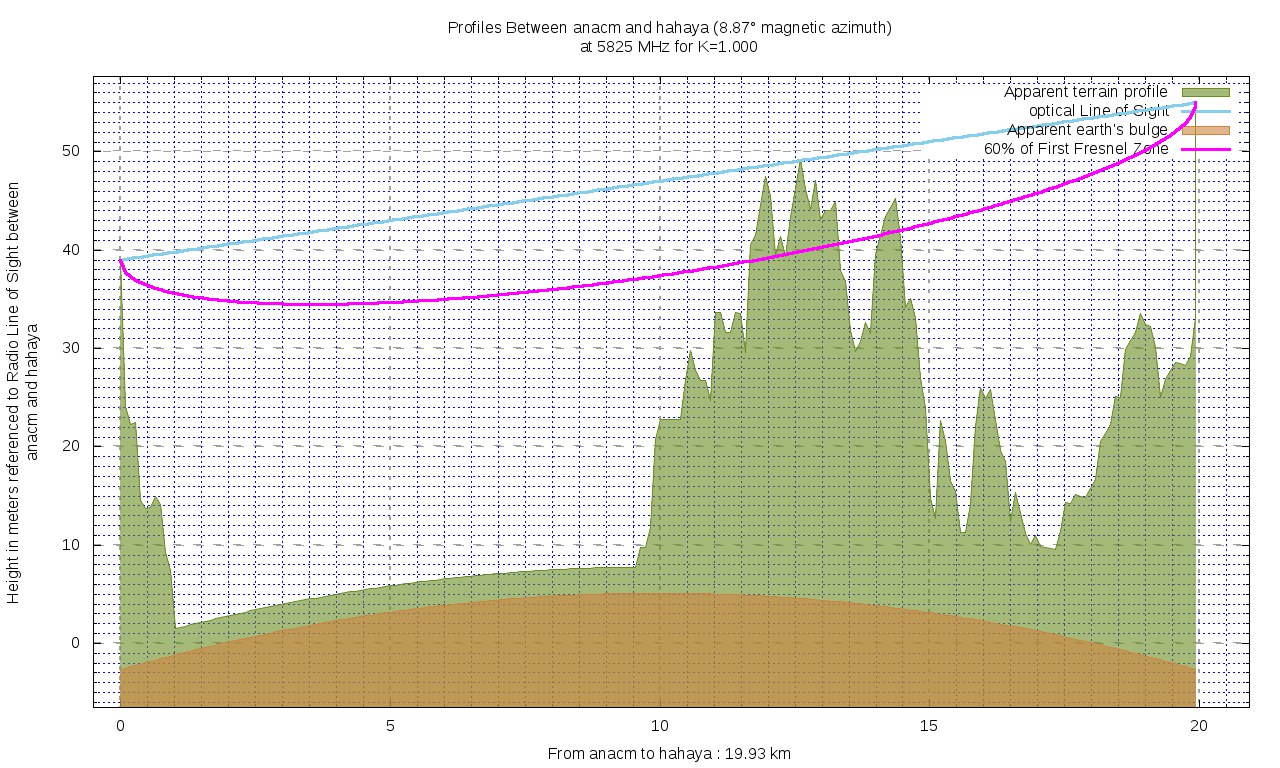}
\caption{Grazing condition: Blue line is the optical LOS, magenta is the Fresnel zone lower contour, green the terrain profile and brown the earth curvature. This link is partially obstructed with a 6 dB loss as a consequence.}
\label{2}
\end{figure}

In conclusion, for best results we want to clear at least 60\% of the Fresnel zone, and use enough antenna gain to make sure that the received signal is greater than the receiver sensitivity. The difference is the margin of the link, and the greater the margin the more reliable the link will be. The receiver sensitivity is specified by the vendor of the equipment for different transmission speeds, and it is determined by the electrical noise and the required signal to noise ratio to reliably decode the information. The greater the speed, the less sensitive the receiver will be. This means that if the link proves infeasible at say, 100 Mbps, we might still operate it if we set the capacity to 25 Mbps. This tradeoff between speed and sensitivity is present in any communication system and is due to the fact that we need a certain amount of energy to detect the received signal in the presence of the inevitable noise, when we increase the transmission speed  we are sending more symbols in the same time and therefore the energy per symbol will decrease and so will  the signal to noise ratio.

\section{Telegram and Telegram Bots}

Telegram \cite{teg} is a cloud-based instant messaging service over which any type of file can be exchanged. Telegram clients exist for smartphones (Android, iOS, Windows Phone, Ubuntu Touch) and desktop systems (Windows, OS X, Linux). It differs from whatsapp \cite{wat} in that it will not support voice calls, but unlike whatsapp it can be used without a phone. The client code is open software, but the source is proprietary. It is free and has no limits on the size of media. Servers are distributed in many locations around the world.
In June 2015, Telegram launched a platform for third-party developers to create bots. Bots are Telegram accounts operated by programs that respond to messages or mentions and can be integrated in other programs. They mimic the behaviour of a human being in specific applications, like a help desk.
\

\section{BotRf usage Example}
The head of the Telecommunications unit at Universidad de los Andes in Merida, Venezuela, asked one of his staff to use BotRf as a guide for the planning on wireless links that were needed to extend the university network from the administrative building to the printing shop. There are many buildings that block the line of sight between the end points so a two legs solution had to be devised.
Plan del Morro was chosen as repeater site since due to its altitude it offers an unencumbered  view of both ends. Figure 2 shows the layout of the two ends points and the proposed repeater site.
\vspace{100 mm}

\begin{figure}[h]
\centering
\includegraphics[width=4.5in]{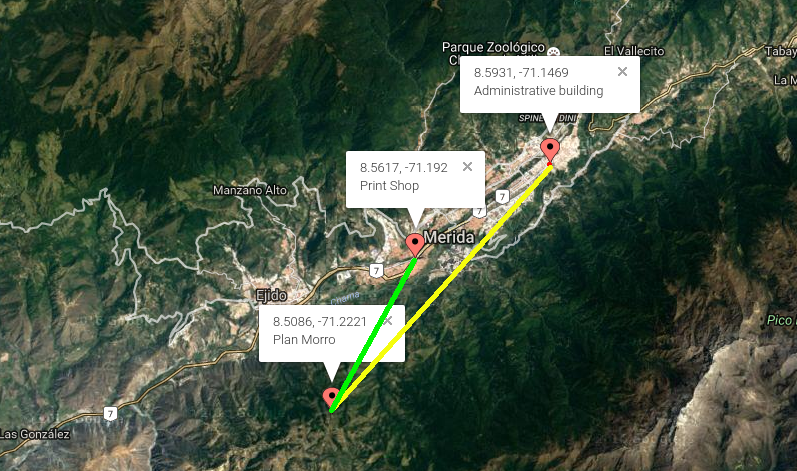}
\caption{Layout of the two legs link from adm to Print Shop, Merida Venezuela.}
\label{3}
\end{figure}
Entering the coordinates and antennas heights in BotRf and using the calc command, figure 3 was shown after few seconds:

\begin{figure}[h]
\centering
\includegraphics[width=4.5in]{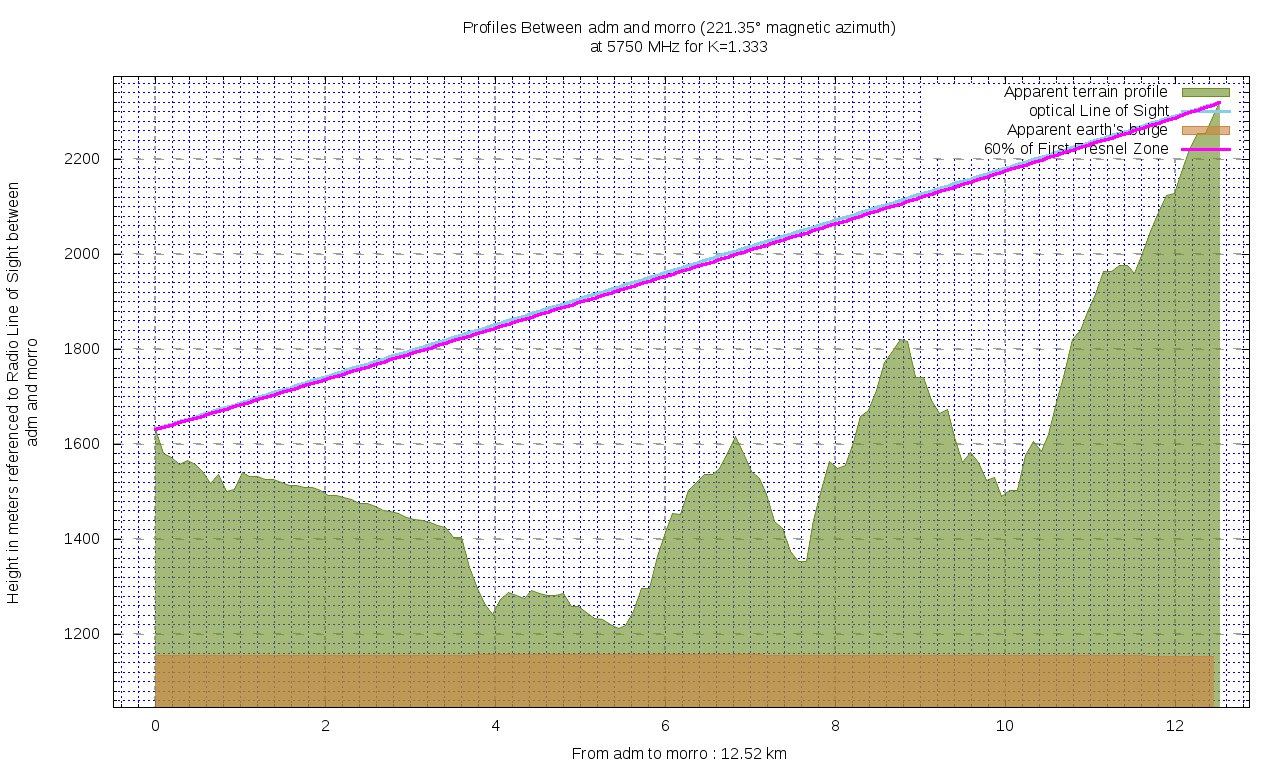}
\caption{Terrain profile between edificio adm and plan del morro, Merida Venezuela.}
\label{3}
\end{figure}

Entering rep in BotRF produced the extensive report:
\vspace{5 mm}

Transmitter site: edif\_ adm
Site location: 8.5931 North / 71.1469 West

Elevation: 1582 m above sea level,
Antenna height: 50 m above ground 

Distance to plan morro: 12.52 km
Azimuth to plan morro: 221.3 degrees

Elevation: +3.1degrees

Receiver site: plan\_ morro

Site location: 8.5086 North / 71.2221 West 

Elevation: 2311 m above sea level,
antenna height: 6 m above ground

Azimuth to edif\_ adm: 41.3 degrees,
depression angle: -3.2 degrees

Free space path loss: 129.69 dB
Longley-Rice path loss: 129.55 dB

Attenuation due to terrain shielding: -0.13 dB

Mode of propagation: Line-Of-Sight 

No obstructions to LOS due to terrain were detected by BotRf:

The first Fresnel zone is clear.
\vspace{5 mm}

Similarly, the link between plan\_ morro and the print shop was simulated with BotRf and the following results were obtained:
Transmitter site: plan\_ morro

Distance to Print Shop: 6.77 km,
azimuth: 29.2 degrees
Depression angle: -8.5 degrees

Receiver site: print shop

Site location: 8.5617 North / 71.1920 West 
Elevation: 1315 m above sea level

Antenna height: 5 m above ground,
distance to plan morro: 6.77 km,
azimuth: 209.3 degrees
Elevation angle: +8.5 degrees

Free space path loss: 124.35 dB
Longley-Rice path loss: 124.18 dB

Atten. due to terrain shielding: -0.17 dB
Mode of propagation: Line-Of-Sight 

\vspace{5 mm}

This proved that both links are feasible and that the connection between Edif Adm and the Print Shop can be accomplished provided that the system gain of the radios and antennas is greater than 130 dB plus the required link margin. 
The BotRf  cnv  command will facilitate the conversion between mW and dB, between dBuV/m and dBm, between frequency and wavelength, etc.
The pow command in BotRf is used for to obtain the power levels in the link. The user needs to input the name of the transmitter site, the name of the receiver site, the transmitter power output in dBm, the cable loss between the transmitter and the antenna in dB, the gain of the transmitter antenna in dBi (dB with respect to an isotropic antenna), the gain of the receiver antenna in dBi, the cable loss in dB between the antenna and the receiver and the receiver sensitivity in dBm.

So, in this case we have:

\vspace{5 mm}
pow edif\_adm plan\_morro 20  0  24 24 0 -87

\vspace{3 mm}
which will generate the graph

\vspace{80 mm}

\begin{figure}[h]
\centering
\includegraphics[width=4.5in]{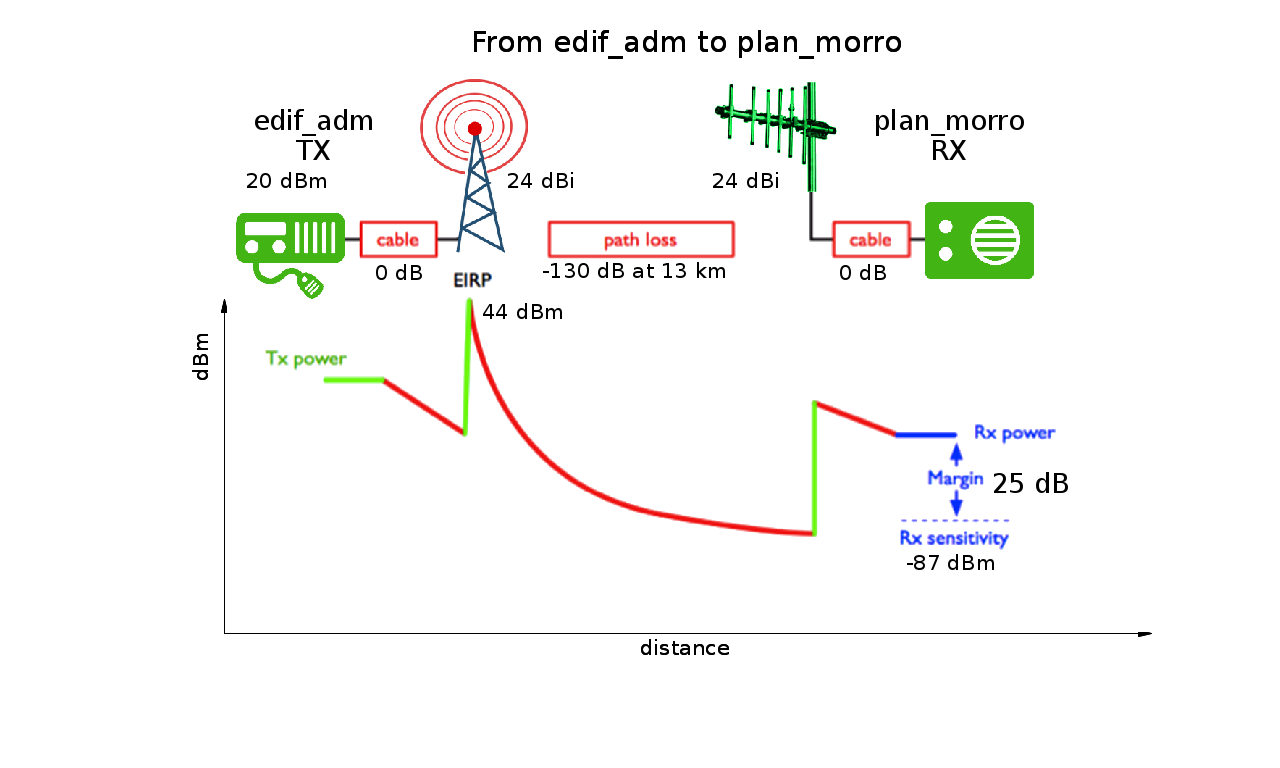}
\caption{Power versus distance along the link between edif\_adm and plan\_morro.}
\label{2}
\end{figure}

In which we can see the values of the EIRP (Equivalent Isotropic Radiated Power), as well as the link margin, 25 dB, providing a quite reliable signal that will support the highest modulation schemes. The path loss used is derived directly from the free space loss previously calculated by BotRf. The additional attenuation that might be present must be considered in each case and added to the free space loss to obtain the new  margin.

It is worth noting that all the input data, as well as the results, are stored in the server under each user's account, so they can be retrieved for future use.  BotRf list command will provide a list of all the sites ever entered with their respective coordinates

\section{Conclusions and future works}
We have presented a user friendly and platform independent tool that can be used for the planning of RF links even by people of limited telecommunications background. BotRf can also be used as a learning tool to asses the impact of the different elements of the communication channel in the final performance. In particular, the apparent radius of curvature of the earth can be modified to determine the effect of the changing of the refractive index. The  terrain profile is derived from publicly available digital elevation maps of high resolution that are housed on the server where all the calculations are made in real time, whereby the Internet traffic used is at a minimum, therefore  the tool can be used in places with very basic telecommunications infrastructure. Two different models can be invoked to asses the additional attenuation introduced by the effect of the terrain. We are currently working on the point to area coverage prediction and into the implementation of other RF propagation models and also in the use  of the tool for the planning of a TV white Spaces deployment in Mozambique. The user interface will soon be available in Spanish and French.


%

\begin{thebibliography}{5}
\bibitem{wndw} Wireless Networking in the Developing World, third edition (2013) http://wndw.net 

\bibitem{srtm} Shuttle Radar Topography Mission (SRTM) 1 Arc-Second Global https://lta.cr.usgs.gov/SRTM1Arc



\bibitem{rm} Radio Mobile Home Page, http://www.cplus.org/rmw/english1.html

\bibitem{rmol} Radio Mobile On-Line Home Page, http://www.cplus.org/rmw/rmonline.html

\bibitem{qsl} Magliacane  J.  SPLAT!  An  RF  Signal Propagation, Loss, And Terrain analysis tool for the spectrum between 20 MHz and 20 GHz http://www.qsl.net/kd2bd/splat.htm 


\bibitem{wat} https://www.whatsapp.com/
\bibitem{bot} https://en.wikipedia.org/wiki/Interne
\bibitem{cal} Caleb Phillips, Douglas Sicker, and Dirk Grunwald,?Bounding the Practical Error of Path Loss Models?, in Hindawi Publishing Corporation International Journal of Antennas and Propagation, Volume 2012, Article ID 754158, 21 pages, 2012.
\bibitem{longl} Longley, A. G., and P. L. Rice , Prediction of tropospheric radio transmission loss over irregular terrain, ESSA Tech. Report ERL 79-ITS 67 (1968).
\bibitem{itu} Propagation data and prediction methods required for the design of terrestrial line-of-sight systems, ITU-R P.530, 2015
\bibitem{hey} http://www.heywhatsthat.com/profiler.html
\bibitem{qrp}  http://qradiopredict.sourceforge.net/
\bibitem {crc} http://lrcov.crc.ca/main/
\bibitem{piast} http://piast.edu.pl/Sites/subsite\_TerrainProfileVisualization.aspx 
\bibitem{teg} https://telegram.org
\end{thebibliography}
\end{document}